\documentclass[9pt,twocolumn,twoside]{osajnl}

\journal{josab} 

\setboolean{shortarticle}{false}
\usepackage{braket}

\def\bcen{\begin{center}}
\def\ecen{\end{center}}
\usepackage{epstopdf}
\usepackage{mathtools, cuted}
\usepackage{lipsum, color}
\ifthenelse{\boolean{shortarticle}}{\colorlet{color2}{color2b}}{\colorlet{color2}{color2}} 

\title{Dependence of the photon statistics of down-converted field-modes on the photon statistics of pump field-mode}

\author[1,*]{Nilakantha Meher}
\author[1,$\dagger$]{Anand K. Jha}%

\affil[1]{Department of Physics, Indian Institute of Technology Kanpur, Kanpur, UP 208016, India.}
\affil[*]{nilakantha.meher6@gmail.com}
\affil[$\dagger$]{akjha9@gmail.com}

\dates{Compiled \today}

\ociscodes{ (270.0270) Quantum Optics (270.5290)   Photon statistics}

\doi{\url{http://dx.doi.org/10.1364/ao.XX.XXXXXX}}

\begin{abstract}
Using the zero time-delay second-order correlation function for studying the photon statistics, we investigate how the photon statistics of the field-modes generated by parametric down-conversion (PDC) process depends on the photon statistics of the pump field-mode. We derive general expressions for the zero time-delay second-order correlation function of the down-converted field-modes for both multi-mode and single-mode PDC processes. We further study these expressions in the weak down-conversion limit. We show that for a two-photon two-mode PDC process, in which a pump photon splits into two photons into two separate field-modes, the zero time-delay second-order correlation function of the individual down-converted field-modes is equal to twice that of the pump field-mode. Furthermore, for an $n$-photon $n$-mode down-conversion process, in which a pump photon splits into $n$ photons into $n$ separate field-modes, the zero time-delay second-order correlation function of the individual down-converted field-modes is equal to $2^{(n-1)}$ times that of the pump field-mode. However, in contrast to the multi-mode PDC processes, for a single-mode PDC process, in which a pump photons splits into two or more photons into a single mode, the zero time-delay second-order correlation function of the down-converted field-mode is not proportional to that of the pump in the weak down-conversion limit. Nevertheless, we find it to be inversely proportional to the average number of photons in the pump field-mode.  
\end{abstract}

\setboolean{displaycopyright}{true}

\begin{document}

\maketitle
\thispagestyle{fancy}

\ifthenelse{\boolean{shortarticle}}{\ifthenelse{\boolean{singlecolumn}}{\abscontentformatted}{\abscontent}}{}
\section{Introduction}
Parametric down-conversion (PDC) is a nonlinear
process in which a pump photon of higher frequency splits into two or more photons of lower frequencies \cite{Louisell1961PR, Walls1970PRA, Hillery1985PRA,Drobny1993PRA,Drobny1994PRA, Drumond2014Cambridge, Corona2011OptLett, Douady2004OptLett,Chang2020PRX}. In case of two-photon PDC, the down-converted photons are called the signal and idler photons \cite{Burnham1970PRL}. Generation of these photons satisfies both energy and momentum conservation laws, that is, the sum of the energies of the signal and idler photons is equal to the energy of the pump photon and the sum of the momenta of the signal and idler photons is equal to the momentum of the pump photon.  Due to energy and momentum conservations, these down-converted photons become entangled in various degrees of freedom such as energy-time \cite{Kwiat1993PRA,Strekalov1996PRA}, polarization \cite{Rubin1994PRA, Kwiat1994PRA}, position-momentum \cite{Howell2004PRL}, angular momentum-angle \cite{Rarity1990PRL, Leach2010Science}, etc. The entanglement of PDC photons is used in various applications such as quantum teleportation \cite{Bouwmeester1997Nature}, quantum gates \cite{OBrien2003Nature}, quantum cryptography \cite{Jennewein2000PRL}, etc. 

Many previous studies have focused on the characterization of down-converted field-modes in various degrees of freedom \cite{Hong1985PRA, Rubin1996PRA, Monken1998PRA,Jha2010PRA,Jha2008PRA,Meher2020JOSAB,Kulkarni2016PRA, Mollow1967PRA, Hillery1984PRA,Crouch1988PRA,Kinsler1993PRA, Hillery1994PRA,Cohen1995PRA, Drobny1993PRA, Olsen2002OptComm}. Several of these studies are based on the parametric approximation, wherein the pump field-mode is considered as a strong classical field-mode and the operators associated with the pump field-mode are replaced by complex numbers in  the PDC Hamiltonian \cite{Mollow1967PRA, Hillery1984PRA,Crouch1988PRA,Kinsler1993PRA}. On the other hand, treating the pump as a quantized field-mode opens up the possibility of studying many other interesting phenomena such as squeezing \cite{Hillery1994PRA,Cohen1995PRA}, photon number correlation \cite{Drobny1993PRA, Olsen2002OptComm}, phase correlation \cite{Gantsog1991OptComm}, sub-Poissonian photon statistics \cite{Bandilla1995PRL}, non-classical states generation \cite{Hillery1994PRA, Niu2017PRL}, etc. 

In this paper, we treat the pump as a quantized field-mode and investigate how the photon statistics of the pump field-mode affects the statistics of the down-converted field-modes.  As the PDC is a unitary process, the state of the down-converted field-modes is calculated by solving the unitary time-dynamics of the pump field-mode governed by the Hamiltonian that describes the PDC process \cite{Bandilla1996PRA, Drobny1993PRA, Vintskevich2019PRA, Bandilla2000JOptB}. We characterize the photon statistics of the pump and the down-converted field-modes using the second-order correlation function, which is defined as  \cite{Glauber1963PR,Mandel1995,Loudon1983}
\begin{align}
g^{(2)}(\tau)=\frac{\langle a^\dagger (0)a^\dagger (\tau)a(\tau)a(0)\rangle}{\langle a^\dagger (0)a(0)\rangle^2},
\end{align}
where $a^\dagger (a)$ is the creation (annihilation) operator of the field-mode. When the time delay $\tau=0$, the corresponding correlation function $g^{(2)}(0)$ is referred to as the zero time-delay second-order correlation function. It is a very important quantity and is used for studying the quantum aspects of light such as anti-bunching and sub-Poissonian statistics \cite{Kimble1977PRL,Zou1990PRA,Davidovich1996RMP, Morgan1966PRL,Eisaman2011RSI,Short1983PRL}. Moreover, $g^{(2)}(0)$ is also used as a measure of purity for single photon sources \cite{Bocquillon2009PRA,Faruque2019PRApplied,Christ2011NJP}. A field-mode with $g^{(2)}(0)$ larger than one shows super-Poissonian photon statistics whereas a field-mode with $g^{(2)}(0)$ smaller than one shows sub-Poissonian photon statistics. For a Poissonian field-mode, $g^{(2)}(0)$ is unity. Using the zero time-delay second-order correlation functions, we investigate how the photon statistics of the field-modes generated by multi-mode and single-mode parametric down-conversion processes depends on the photon statistics of the pump field-mode. We derive general expressions for the $g^{(2)}(0)$ of the down-converted field-modes and further study them under weak down-conversion limit.

This article is organized as follows: In Sec. \ref{NonDegPDC}, we   consider the multi-mode PDC processes and derive relations between the zero time-delay second-order correlation function of the pump field-mode with that of the individual down-converted field-modes. In Sec. \ref{degeneratePDC}, we investigate the photon statistics in the single-mode PDC process.  Finally, we summarize our results in Sec. \ref{Summary}.
\section{Multi-mode PDC process}\label{NonDegPDC}
In this section, we discuss multi-mode PDC processes, in which a pump photon splits into two or more photons into as many separate field-modes \cite{Drummond1990PRA,Fernee1995PRA,Schneeloch2019JOpt,Guo2017LSA,Paleari2004OptExp,Couteau2018ContempPhy}.
\subsection{Two-photon two-mode PDC process}
First, we consider the two-photon two-mode PDC process, in which a pump photon splits into two photons into two separate modes. The Hamiltonian for the two-photon two-mode PDC process is given by \cite{Drobny1993PRA,Drumond2014Cambridge, McNeil1983PRA, Agrawal1974JPhyA}
\begin{align}\label{Hamiltonian_SchrPict}
H=\omega_p a_p^\dagger a_p+ & \omega_s a_s^\dagger a_s+\omega_i a_i^\dagger a_i \nonumber\\
&+\eta(a_{p}a_{s}^\dagger a_{i}^\dagger+a_{p}^\dagger a_{s}a_{i}),
\end{align}
where $\omega_p,\omega_s,$ and $\omega_i$ are the frequencies of the pump, signal and idler field-modes respectively. They satisfy $\omega_p=\omega_s+\omega_i$. The field-mode operators $a_p(a_p^\dagger), a_s(a_s^\dagger)$ and $a_i(a_i^\dagger)$ are the annihilation (creation) operators for the pump, signal and idler field-modes, respectively.  The coupling constant $\eta$ is given by \cite{Schneeloch2019JOpt}
\begin{align}
\eta\approx\frac{\sigma_p^2}{\sigma_1^2+2\sigma_p^2}\sqrt{\frac{16 \hbar \pi^3 c^3 \chi_{eff}^{(2)}}{\epsilon_0 \mu_s^2 \mu_i^2 \mu_p^2 L \lambda_p^3 \sigma_p^2}},
\end{align}
where $\chi_{eff}^{(2)}$ is the effective second-order nonlinearity. $\sigma_1$ is the field-mode diameter of the signal-idler detection system \cite{Schneeloch2019JOpt} and $\sigma_p$ is the pump beam waist. The refractive index of the pump, signal and idler photons inside the crystal is given by $\mu_p, \mu_s$ and $\mu_i$ respectively. $L$ is the length of the crystal, $\lambda_p$ is the wavelength of the pump, and $c$ is the speed of light in free space.

The Hamiltonian given in Eq.~(\ref{Hamiltonian_SchrPict}) can be written in the interaction picture as \cite{Drobny1993PRA,Drumond2014Cambridge, McNeil1983PRA, Agrawal1974JPhyA}
\begin{align}\label{Hamiltonian_Nondegenerate}
H_I=\eta(a_{p}a_{s}^\dagger a_{i}^\dagger+a_{p}^\dagger a_{s}a_{i}).
\end{align}
In most of the studies, the above interaction Hamiltonian is solved by considering parametric approximation, that is, by replacing the pump annihilation and creation operators by  complex numbers \cite{Mollow1967PRA, Hillery1984PRA,Crouch1988PRA,Kinsler1993PRA,Drumond2014Cambridge}.   In this approximation, $H_I$ becomes $\eta(\alpha_p a_{s}^\dagger a_{i}^\dagger+\alpha_{p}^* a_{s}a_{i})$, where $a_p$ and $a_p^\dagger$ are replaced by complex numbers $\alpha_p$ and $\alpha_p^*$ respectively. However, this is true only if the pump field-mode is in the coherent state. But, the pump field-mode operators cannot be replaced by complex numbers if the pump field-mode is anything other than the coherent state. 

 In this study, we consider the Hamiltonian given in Eq.~(\ref{Hamiltonian_Nondegenerate}).  Let the initial state of the pump be $\ket{\psi}_p$, and the signal and idler field-modes be in their respective vacuua. Thus, at time $t=0$, the state $\ket{\Psi(0)}$ of the pump, signal and idler field-modes can be written as $\ket{\Psi(0)}=\ket{\psi}_p\ket{0}_s\ket{0}_i$, where $\ket{\psi}_p=\sum_{k}c_{k}\ket{k}_p$ and $c_k$ satisfies $\sum_{k}|c_k|^2=1$. Here, $\ket{m}_{p}$ represents the state of the pump field-mode with $m$ photons in it, etc. The state produced by the PDC process at time $t$ can be shown to be
\begin{align}\label{SPDCState}
&\ket{\Psi(t)}=e^{-iH_It}\ket{\Psi(0)},\nonumber\\
 =&(I-itH_{I}-\frac{t^2}{2}H_{I}^2+\frac{it^3}{3!}H_I^3+\frac{t^4}{4!}H_I^4-..)\ket{\psi}_p\ket{0}_s\ket{0}_i,\nonumber\\
=&\frac{1}{\sqrt{N_{1}}}[A_0\ket{\psi}_p\ket{0}_s\ket{0}_i+A_1\ket{\psi}_p\ket{1}_s\ket{1}_i\nonumber\\
+& A_2\ket{\psi}_p\ket{2}_s\ket{2}_i+ A_3\ket{\psi}_p\ket{3}_s\ket{3}_i+A_4\ket{\psi}_p\ket{4}_s\ket{4}_i+..], 
\end{align}
where
\begin{subequations}\label{Coefficients}
\begin{align}
A_0 &=1-\left(\frac{\eta^2 t^2}{2!}-\frac{\eta^4 t^4}{4!}\right) a_p^\dagger a_p+\frac{5\eta^4 t^4}{4!}a_p^{\dagger 2}a_p^2+..,\\
A_1 &=\left(-i\eta t+\frac{i\eta^3 t^3}{3!}\right) a_p+\frac{5i\eta^3 t^3}{3!}a_p^\dagger a_p^2+... ,\\
A_2 &=\left(-\eta^2 t^2+\frac{1}{2}\eta^4 t^4\right) a_p^2+\frac{7\eta^4t^4}{6}a_p^\dagger a_p^3+..,\\
A_3 &= i\eta^3 t^3 a_p^3+..,\\
A_4 &=\eta^4 t^4 a_p^4+...,
\end{align}
\end{subequations}
and $N_1=\sum_{j=0}\langle A_j^\dagger A_j\rangle$ is the normalization constant. Here $\langle A_j^\dagger A_j \rangle=~_p\langle \psi|A_j^\dagger A_j|\psi\rangle_p$. The interaction time $t$ is defined by the length of the crystal \cite{Vintskevich2019PRA}. The state of the signal  and idler field-modes can be calculated from $\rho(t)=\ket{\Psi(t)}\bra{\Psi(t)}$ by tracing over the pump field-mode. Thus, the reduced density matrix $\rho_{si}(t)$ of the signal and idler field-modes in the basis $\{\ket{0}_s\ket{0}_i,\ket{1}_s\ket{1}_i,\ket{2}_s\ket{2}_i,\ket{3}_s\ket{3}_i, \ket{4}_s\ket{4}_i,...\}$ can be written as 
\begin{align}\label{signalidler2}
\rho_{si}(t)&=\text{Tr}_{p} (\ket{\Psi(t)}\bra{\Psi(t)}),\nonumber\\
=\frac{1}{N_1}&\left[\begin{array}{cccccc}
\langle A_0^\dagger A_0\rangle & \langle A_0^\dagger A_1\rangle & \langle A_0^\dagger A_2\rangle & \langle A_0^\dagger A_3\rangle & \langle A_0^\dagger A_4\rangle & \cdots\\
\langle A_1^\dagger A_0\rangle & \langle A_1^\dagger A_1\rangle & \langle A_1^\dagger A_2\rangle & \langle A_1^\dagger A_3\rangle & \langle A_1^\dagger A_4\rangle & \cdots\\
\langle A_2^\dagger A_0\rangle & \langle A_2^\dagger A_1\rangle & \langle A_2^\dagger A_2\rangle & \langle A_2^\dagger A_3\rangle & \langle A_2^\dagger A_4\rangle & \cdots\\
\langle A_3^\dagger A_0\rangle & \langle A_3^\dagger A_1\rangle & \langle A_3^\dagger A_2\rangle & \langle A_3^\dagger A_3\rangle & \langle A_3^\dagger A_4\rangle & \cdots\\
\langle A_4^\dagger A_0\rangle & \langle A_4^\dagger A_1\rangle & \langle A_4^\dagger A_2\rangle & \langle A_4^\dagger A_3\rangle & \langle A_4^\dagger A_4\rangle & \cdots\\
\vdots & \vdots &\vdots & \vdots & \vdots &\ddots
\end{array}\right].
\end{align}
The element $\langle A_m^\dagger A_m\rangle/N_1$ is the  probability of detecting $m$ photons each in the signal and idler field-modes.
Now, the state of the signal field-mode can be obtained from $\rho_{si}(t)$ by tracing it over the idler field-mode. Thus, the state of the signal field-mode in the basis $\{\ket{0}_s,\ket{1}_s,\ket{2}_s,\ket{3}_s, \ket{4}_s,...\}$ can be written as
\begin{align}\label{signalState2}
\rho_{s}(t)&=\text{Tr}_{i} (\rho_{si}),\nonumber\\
=\frac{1}{N_1}&\left[\begin{array}{cccccc}
\langle A_0^\dagger A_0\rangle & 0 & 0 & 0 & 0 & \cdots\\
0 & \langle A_1^\dagger A_1\rangle & 0 & 0 & 0 & \cdots\\
0 & 0 & \langle A_2^\dagger A_2\rangle & 0 & 0 & \cdots\\
0 & 0 & 0 & \langle A_3^\dagger A_3\rangle & 0 & \cdots\\
0 & 0 & 0 & 0 & \langle A_4^\dagger A_4\rangle & \cdots\\
\vdots & \vdots &\vdots & \vdots & \vdots &\ddots
\end{array}\right].
\end{align}
A similar expression can be obtained for $\rho_{i}$ by taking the partial trace of $\rho_{si}$ over the signal field-mode. We note that $\rho_{s}=\rho_{i}$ and therefore, we present our analysis in the subsequent discussions only for the signal field-mode. 

Now, we discuss how the photon statistics of the signal and idler field-modes depend on the photon statistics of the pump field-mode. The zero time-delay second-order correlation function of the signal field-mode is given by
$g^{(2)}_s(0)=\langle a_s^{\dagger 2}a_s^2\rangle /\langle a_s^\dagger a_s\rangle^2$,
where $\langle a_s^{\dagger 2}a_s^2 \rangle=\text{Tr}(\rho_s a_s^{\dagger 2}a_s^2)$ and $\langle a_s^{\dagger }a_s \rangle=\text{Tr}(\rho_s a_s^{\dagger }a_s)$. 
Using the state given in Eq.~(\ref{signalState2}), we find 
\begin{align}\label{g2sA}
g_s^{(2)}(0)=\frac{N_1\sum_{k=2} k(k-1)\langle A_k^\dagger A_k \rangle}{\left(\sum_{k=1} k\langle A_k^{\dagger} A_k \rangle \right)^2}.
\end{align}
By using the operators given in Eqs.~(\ref{Coefficients}$a$-\ref{Coefficients}$e)$, we write Eq.~(\ref{g2sA}) to be
\begin{strip}
\vspace{1.5cm}
\begin{align}\label{gpgsgi}
g^{(2)}_s(0)=2g_p^{(2)}(0)\frac{N_1[1-\eta^2 t^2+\frac{2}{3}\eta^2 t^2 \frac{g_p^{(3)}(0)}{g_p^{(2)}(0)}n_p+....]}{\left[1-\frac{1}{3}\eta^2 t^2+\frac{1}{36}\eta^4 t^4 +\left(\frac{1}{3}\eta^2 t^2-\frac{37}{36}\eta^4 t^4\right)g_p^{(2)}(0)n_p-\frac{35}{36}\eta^4 t^4 g_p^{(3)}(0)n_p^2+.....\right]^2},
\end{align}
\end{strip}
where $g_p^{(k)}(0)=\langle a_p^{\dagger k}a_p^k\rangle/\langle a_p^\dagger a_p \rangle^k$ is the $k$th order correlation function and $n_p=\langle a_p^\dagger a_p\rangle$ is the average number of photons in the pump field-mode. As $\rho_s=\rho_i$, we get $g^{(2)}_i(0)=g^{(2)}_s(0)$.  

Equation~(\ref{gpgsgi}) shows the dependence of the photon statistics of individual down-converted field-modes on the photon statistics of the pump field-mode. It is valid for arbitrary pump field-mode states with any down-conversion strength $n_p \eta^2 t^2$. A wide range of down-conversion strength is achievable experimentally using continuous-wave laser in non-degenerate optical parametric amplifier \cite{Wang2010OptExp,Yan2012PRA,Zhou2015OptExp,Vintskevich2019PRA}. However, many experiments fall under the category of being in the weak down-conversion limit, that is, $n_p \eta^2 t^2\ll 1$. For instance, a 404 nm pump with 100mW power having pump radius of 0.4 mm incident on a 3 mm long BiBO crystal gives $\eta\sim 2.85\times 10^3$ \cite{Schneeloch2019JOpt}. The average number of pump photons inside the crystal at any given time is $n_p=\frac{P}{\hbar \omega}\frac{L\mu_p}{c}=3.7\times 10^6$, where $P$ is the pump power and $\mu_p$ is the refractive index of the pump inside the crystal. The time $t$, which is the traveling time of the pump in the crystal,  is about $\sim 10^{-11}$sec. This gives the down-conversion strength $n_p\eta^2 t^2\sim 10^{-10}$. In the weak down-conversion limit $n_p\eta^2 t^2 \ll 1$, Eq.~(\ref{gpgsgi}) can be approximated as
\begin{align}\label{g22gp}
g^{(2)}_s(0)\simeq 2g_p^{(2)}(0).     
\end{align}
This comes from the fact that the numerator and the denominator in Eq.~(\ref{gpgsgi}) become unity in the limit $n_p \eta^2 t^2\ll 1$. 
A similar expression can be found for the idler field-mode, that is, $g^{(2)}_i(0)\simeq 2g_p^{(2)}(0)$. The signal and idler field-modes become super-Poissonian if $g_p^{(2)}(0)>1/2$,  and remain sub-Poissonian if $g_p^{(2)}(0)<1/2$. Hence, in order to produce sub-Poissonian signal and idler field-modes, the zero time-delay second-order correlation function of the pump field-mode has to be less than 1/2.
Moreover, the pump field-mode with $g_p^{(2)}(0)=1/2$ produces signal and idler field-modes with Poissonian photon distributions. As these down-converted field-modes are mixed individually (refer Eq.~(\ref{signalState2})), they are known as mixed Poissonian states \cite{Malbouisson2000PhysicaA}. Also, Eq.~(\ref{g22gp}) immediately recovers a known result that the photon statistics of the signal  and idler field-modes are thermal $(g^{(2)}_s(0)=g^{(2)}_i(0)=2)$ if the pump field-mode is a coherent state $(g_p^{(2)}(0)=1)$ \cite{Mollow1967PRA,Drobny1993PRA,Paleari2004OptExp}.
\subsection{n-photon n-mode PDC process}
Next, we consider an $n$-photon down-conversion process, in which a pump photon splits into $n$ photons of lower frequencies. These $n$ down-converted photons go into $n$ separate modes, and hence, the interaction Hamiltonian for the $n$-photon $n$-mode down-conversion process is given by \cite{Hillery1985PRA,Drobny1993PRA,Drobny1994PRA, Drumond2014Cambridge}
\begin{align}\label{Hamiltonian_NondegenerateN}
H_{I}=\eta ( a_{p} \otimes_{j=1}^{n} a_{j}^\dagger +a_{p}^\dagger \otimes_{j=1}^{n} a_{j}),
\end{align}
where $\otimes_{j=1}^{n} a_j=a_1\otimes a_2 \otimes ...\otimes a_n$. Here, $a_j$ is the annihilation operator corresponds to $j$th down-converted field-mode. The frequencies of these field-modes satisfy $\omega_p=\omega_1+\omega_2+...+\omega_n$. Here, $\omega_p$ is the frequency of the photons in the pump field-mode mode and $\omega_j$ is the frequency of the photon in the $j$th down-converted field-mode.

Let us consider that the initial state $\ket{\Psi(0)}$ of the pump and the down-converted field-modes at $t=0$ is $\ket{\Psi(0)}=\ket{\psi_p}\ket{0}_1\ket{0}_2...\ket{0}_n$, where $\ket{0}_j$ is the vacuum state corresponds to the $j$th down-converted field-mode. The state of the down-converted field-modes can be obtained by solving the unitary dynamics governed by the Hamiltonian given in Eq.~(\ref{Hamiltonian_NondegenerateN}) (refer Appendix. \ref{AppendixA} for calculation). From the evolved state, we find the zero time-delay second-order correlation function for the $j$th down-converted field-mode to be 
\begin{strip}
\begin{align}\label{gjn}
g_j^{(2)}(0)=2^{(n-1)}g_p^{(2)}(0)\frac{N_2\left[1-\frac{(2+2^n)}{6}\eta^2 t^2-\left(\frac{(1+2^n)2^{n/2}+(3!)^{n/2}3^{n/2}}{2^{n/2}3!}-\frac{2(3!)^n}{2^{n}3!}\right)n_p\eta^2 t^2\frac{g_p^{(3)}(0)}{g_p^{(2)}(0)}+...\right]}{\left[1-\frac{1}{3}\eta^2 t^2+\frac{1}{36}\eta^4 t^4+\left(2^{(n-1)}-\frac{(1+2^n)}{3}\right)n_p \eta^2 t^2 g_p^{(2)}(0)+..... \right]}.
\end{align}
\end{strip}
This is a general result which shows the dependence of photon statistics of individual down-converted field-modes on the photon statistics of the pump field-mode in $n$-photon $n$-mode down-conversion process.
Now, in the limit $n_p\eta^2 t^2 \ll 1$, the denominator and numerator of Eq.~\ref{gjn} approach unity and the zero time-delay second-order correlation function becomes
\begin{align}\label{gkgpmain}
g^{(2)}_j(0)\simeq 2^{(n-1)}g_p^{(2)}(0). 
\end{align}
This relation is true for all $j$, \textit{i.e.,} for all the down-converted field-modes.  For $n=2$, this result goes over to the result for two-photon two-mode PDC process given in Eq.~(\ref{g22gp}).
Pump field-mode with $g_p^{(2)}(0)=1/2^{(n-1)}$ produces down-converted field-modes with Poissonian photon distributions. These field-modes are the Poissonian mixed states \cite{Malbouisson2000PhysicaA}. The pump state with $g_p^{(2)}(0)>1/2^{(n-1)}$ produces super-Poissonian down-converted field-modes while in the opposite limit it produces sub-Poissonian down-converted field-modes.

\section{Single-mode PDC process}\label{degeneratePDC}
In this section, we consider single-mode PDC process, in which a pump photon splits into two or more photons in the same field-mode \cite{Cohen1995PRA,Keller1997PRA}
\subsection{Two-photon single-mode PDC process}
The Hamiltonian that describes the two-photon single-mode PDC process, in which a single photon splits into two photons in the same field-mode, is  \cite{Drobny1993PRA,Drumond2014Cambridge}
\begin{align}\label{Hamiltonian_degenerate}
\tilde{H}_{I}=\eta(a_{p}a_d^{\dagger 2} +a_{p}^\dagger a_d^2),
\end{align}
where $a_d (a_d^\dagger)$ is the annihilation (creation) operator of the down-converted field-mode. In this case, $\omega_p=2\omega_d$, where $\omega_d$ is the frequency of the down-converted photons. 

Consider the initial state of the pump field-mode and down-converted field-mode is $\ket{\psi}_p\ket{0}_d$, where $\ket{\psi}_p$ is the state of the pump field-mode and $\ket{0}_d$ represents the down-converted field-mode being in the vacuum state. Then the evolved state under the Hamiltonian given in Eq.~(\ref{Hamiltonian_degenerate}) at time $t$ is
\begin{align}
\ket{\Psi(t)}&=\frac{1}{\sqrt{N_3}}[B_0\ket{\psi}_p\ket{0}_d+B_1\ket{\psi}_p\ket{2}_d\nonumber\\
+&B_2\ket{\psi}_p\ket{4}_d+B_3\ket{\psi}_p\ket{6}_d+B_4\ket{\psi}_p\ket{8}_d+....],
\end{align} 
where
\begin{subequations}\label{CoefficientSingle}
\begin{align}
B_0 &=1-\left({\eta^2 t^2}-\frac{\eta^4 t^4}{6}\right) a_p^\dagger a_p+\frac{7\eta^4 t^4}{6}a_p^{\dagger 2}a_p^2+..,\\
B_1 &=i\sqrt{2}\left[\left(-\eta t+\frac{\eta^3 t^3}{3}\right) a_p+\frac{7\eta^3 t^3}{3}a_p^\dagger a_p^2+...\right],\\
B_2 &=\sqrt{6}\left[\left(-\eta^2 t^2+\frac{4}{3}\eta^4 t^4\right) a_p^2+\frac{11\eta^4t^4}{3}a_p^\dagger a_p^3+..\right],\\
B_3 &=i\sqrt{20} \eta^3 t^3 a_p^3+..,\\
B_4 &=\sqrt{70}\eta^4 t^4 a_p^4+.... 
\end{align}
\end{subequations}
Here $N_3=\sum_{j=0}\langle B_j^\dagger B_j\rangle$ is the normalization constant. It is to be noted that the down-converted field-mode carries even number of photons and the probability of detecting odd number of photons is zero. The state of the down-converted field-mode can be calculated by tracing $\rho(t)=\ket{\Psi(t)}\bra{\Psi(t)}$ over the state of the pump field-mode. Hence, the reduced density matrix of the down-converted field-mode in the even photon number basis $\{\ket{0}_d,\ket{2}_d,\ket{4}_d,\ket{6}_d,...\}$  is
\begin{align}\label{StateCollinear}
\rho_{d}(t)&=\text{Tr}_{p} (\ket{\Psi(t)}\bra{\Psi(t)}),\nonumber\\
=\frac{1}{N_3}&\left[\begin{array}{cccccc}
\langle B_0^\dagger B_0\rangle & \langle B_0^\dagger B_1\rangle & \langle B_0^\dagger B_2\rangle & \langle B_0^\dagger B_3\rangle & \langle B_0^\dagger B_4\rangle & \cdots\\
\langle B_1^\dagger B_0\rangle & \langle B_1^\dagger B_1\rangle & \langle B_1^\dagger B_2\rangle & \langle B_1^\dagger B_3\rangle & \langle B_1^\dagger B_4\rangle & \cdots\\
\langle B_2^\dagger B_0\rangle & \langle B_2^\dagger B_1\rangle & \langle B_2^\dagger B_2\rangle & \langle B_2^\dagger B_3\rangle & \langle B_2^\dagger B_4\rangle & \cdots\\
\langle B_3^\dagger B_0\rangle & \langle B_3^\dagger B_1\rangle & \langle B_3^\dagger B_2\rangle & \langle B_3^\dagger B_3\rangle & \langle B_3^\dagger B_4\rangle & \cdots\\
\langle B_4^\dagger B_0\rangle & \langle B_4^\dagger B_1\rangle & \langle B_4^\dagger B_2\rangle & \langle B_4^\dagger B_3\rangle & \langle B_4^\dagger B_4\rangle & \cdots\\
\vdots & \vdots &\vdots & \vdots & \vdots &\ddots
\end{array}\right].
\end{align}
Here $\langle B_j^\dagger B_k \rangle=~_p\bra{\psi} B_j^\dagger B_k \ket{\psi}_p$. The diagonal element, for instance, $\langle B_k^\dagger B_k\rangle/N_3$ gives the probability of detecting $2k$ photons in the down-converted field-mode.

We calculate the zero time-delay second-order correlation function corresponding to the state given in Eq.~(\ref{StateCollinear}) to be
\begin{align}
g_d^{(2)}(0)=\frac{\langle a_d^{\dagger 2}a_d^2\rangle}{\langle a_d^\dagger a_d\rangle^2}=\frac{N_3\sum_{k=1}2k(2k-1)\langle B_k^\dagger B_k\rangle}{\left(\sum_{k=1}2k\langle B_k^\dagger B_k\rangle\right)^2},
\end{align}
where $\langle a_d^{\dagger 2}a_d^2 \rangle=\text{Tr}( a_d^{\dagger 2}a_d^2\rho_{d})$ and $\langle a_d^{\dagger }a_d \rangle=\text{Tr}( a_d^{\dagger }a_d \rho_{d})$. Using the operators given in Eqs.~(\ref{CoefficientSingle}$a$-\ref{CoefficientSingle}$e$), we find
\begin{align}\label{g2bsingle}
g_d^{(2)}(0)=\frac{1}{4n_p\eta^2 t^2}\frac{N_3\left[\left(1-\frac{2}{3}\eta^2 t^2\right)+\frac{40}{3}n_p\eta^2 t^2
g_p^{(2)}(0)+..\right]}{\left[\left(1-\frac{2}{3}\eta^2 t^2\right)+\frac{4}{3}n_p\eta^2 t^2
g_p^{(2)}(0)+..\right]^2},
\end{align}
which shows the dependence of photon statistics of the single-mode down-converted field-mode on the photon statistics of the pump field-mode. 
In the weak down-conversion limit, Eq.~(\ref{g2bsingle}) reduces to
\begin{align}\label{g2dDeg}
g_d^{(2)}(0)\simeq \frac{1}{4 n_p\eta^2 t^2}.
\end{align}
As can be seen, in contrast to the multi-mode PDC process, the $g_d^{(2)}(0)$ of the down-converted field-mode is not proportional to the $g_p^{(2)}(0)$ of the pump field-mode. Nevertheless, we see that the  $g_d^{(2)}(0)$ of the down-converted field-mode is inversely proportional to the average number of pump photons.  As $\eta^2 t^2 n_p \ll 1$, $g_d^{(2)}(0) \gg 1$ and hence, the down-converted field-mode is highly super-Poissonian. 
\subsection{n-photon single-mode PDC process}\label{N-PDC}
The interaction Hamiltonian for this process is \cite{Drobny1994PRA}
\begin{align}\label{Hamiltonian_degenerateN}
\tilde{H}_{I}=\eta(a_{p}a_d^{\dagger n} +a_{p}^\dagger a_d^n),
\end{align}
where $a_d (a_d^\dagger)$ is the annihilation (creation) operator corresponds to the down-converted field-mode. In this case, $\omega_p=n\omega_d$, where $\omega_d$ is the frequency of down-converted photons. 
By considering the state of the pump and down-converted modes at $t=0$ to be $\ket{\psi}_p\ket{0}_d$, where $\ket{\psi}_p$ is the state of the pump field-mode and $\ket{0}_d$ is the vacuum of down-converted field-mode, we calculate the zero time-delay second-order correlation function of the down-converted field-mode at time $t$ to be (refer Appendix. \ref{AppendixB})
\newpage
\begin{strip}
\begin{align}\label{g2singlen}
g_d^{(2)}(0)&=\frac{\langle a_d^{\dagger 2}a_d^2\rangle}{\langle a_d^\dagger a_d\rangle^2}=\frac{n-1}{n! n n_p \eta^2 t^2}\frac{N_4\left[\left(1-\frac{n!}{3}\eta^2 t^2\right)-\left[\frac{1}{3}\left(\frac{(2n)!}{n!}+n!\right)-\frac{2n-1}{2(n-1)}\frac{(2n)!}{n!}\right]n_p\eta^2 t^2
g_p^{(2)}(0)+..\right]}{\left[\left(1-\frac{n!}{3}\eta^2 t^2\right)-\left[\frac{1}{3}\left(\frac{(2n)!}{n!}+n!\right)-\frac{1}{2}\frac{(2n)!}{n!}\right]n_p\eta^2 t^2
g_p^{(2)}(0)+...\right]^2}.
\end{align}
\end{strip}
For $n=2$, which corresponds to the two-photon single mode PDC process, the above expression reduces to the expression given in Eq.~(\ref{g2bsingle}).
In the weak-down-conversion limit, that is, $n_p\eta^2 t^2\ll 1$, the zero time-delay second-order correlation function for $n$-photon single mode PDC process becomes 
\begin{align}
g_d^{(2)}(0)\simeq \frac{n-1}{n! n n_p \eta^2 t^2}.
\end{align}
Hence, the second-order correlation function is inversely proportional to the average number of photons in the pump field-mode. 
\section{Summary}\label{Summary}
We have investigated the role of photon statistics of the pump field-mode in deciding the photon statistics of the down-converted field-modes in multi-mode and single-mode parametric down-conversion processes. We have characterized the photon statistics of the down-converted field-modes in terms of their corresponding second-order correlation functions and have derived general expressions for the zero time-delay second-order correlation function for the down-converted field-modes. In the weak down-conversion limit, we have shown that the values of the second-order correlation functions of the signal and idler field-modes are twice that of the pump field-mode in the two-photon two-mode PDC process.  This result reflects a well known fact that the signal and idler field-modes are thermal if the pump is a coherent state. In general, in the weak down-conversion limit, the zero time-delay second-order correlation function of the individual down-converted field-modes in the $n$-photon $n$-mode down conversion process is equal to $2^{(n-1)}$ times that of the pump field-mode. In contrast to the multi-mode PDC, the zero time-delay second-order correlation function of the single-mode down-converted field-modes is not proportional to that of the pump field-mode in the weak down-conversion limit. Nevertheless, we have found that the zero time-delay second-order correlation function of the single-mode down-converted field-modes is inversely proportional to the average number of photons in the pump field-mode. Although we have used the second-order correlation function $g^{(2)}(0)$ for describing and studying the multi-mode down-converted fields, we note that for a complete description of a field-mode, one has to study the correlation functions of all orders and not just the second-order correlation function  [41]. The higher-order correlation functions may contain very interesting and useful information about the down-converted field studied in this article and may thus become a subject of future research in this direction.
\begin{strip}
\section{Appendix A: n-photon n-mode PDC process}\label{AppendixA}
The interaction Hamiltonian for the $n$-photon $n$-mode down-conversion process is \cite{Hillery1985PRA,Drobny1993PRA,Drobny1994PRA, Drumond2014Cambridge}
\begin{align}
H_{I}=\eta ( a_{p} \otimes_{j=1}^{n} a_{j}^\dagger +a_{p}^\dagger \otimes_{j=1}^{n} a_{j}).
\end{align}
The initial state $\ket{\psi}_p\ket{0}_1\ket{0}_2...\ket{0}_n$ at $t=0$ evolves under the above Hamiltonian as
\begin{align}
\ket{\Psi(t)}&=e^{-iH_It}\ket{\psi}_p\ket{0}_1\ket{0}_2...\ket{0}_n,\nonumber\\
&=\frac{1}{\sqrt{N_2}}[A_0\ket{\psi_p}\ket{0}_1..\ket{0}_n+A_1\ket{\psi_p}\ket{1}_1..\ket{1}_n+A_2\ket{\psi_p}\ket{2}_1..\ket{2}_n+A_3\ket{\psi_p}\ket{3}_1..\ket{3}_n+A_4\ket{\psi_p}\ket{4}_1..\ket{4}_n...,
\end{align}
where
\begin{align*}
A_0 &=1-\left(\frac{\eta^2 t^2}{2!}-\frac{\eta^4 t^4}{4!}\right) a_p^\dagger a_p+\frac{(1+2^n)\eta^4 t^4}{4!}a_p^{\dagger 2}a_p^2+..,\\
A_1 &=\left(-i\eta t+\frac{i\eta^3 t^3}{3!}\right) a_p+\frac{(1+2^n)i\eta^3 t^3}{3!}a_p^\dagger a_p^2+... ,\\
A_2 &=\left(-\frac{2^{n/2}}{2}\eta^2 t^2+\frac{(2+2^n)2^{n/2}}{4!}\eta^4 t^4\right) a_p^2+\frac{(1+2^n)2^{n/2}+(3!)^{n/2}3^{n/2}}{4!}\eta^4t^4a_p^\dagger a_p^3+..,\\
A_3 &= i\frac{(3!)^{n/2}}{3!}\eta^3 t^3 a_p^3+..,\\
A_4 &=\frac{(4!)^{n/2}}{4!}\eta^4 t^4 a_p^4+....
\end{align*}
The state of the $j$th down-converted field-mode can be calculated by tracing over the pump field-mode and rest of the down-converted field-modes except the $j$th field-mode. The reduced density matrix of the $j$th down-converted field-mode is
\begin{align}
\rho_{j}(t)
=\frac{1}{N_2}\left[\begin{array}{cccccc}
\langle A_0^\dagger A_0\rangle & 0 & 0 & 0 & 0 & \cdots\\
0 & \langle A_1^\dagger A_1\rangle & 0 & 0 & 0 & \cdots\\
0 & 0 & \langle A_2^\dagger A_2\rangle & 0 & 0 & \cdots\\
0 & 0 & 0 & \langle A_3^\dagger A_3\rangle & 0 & \cdots\\
0 & 0 & 0 & 0 & \langle A_4^\dagger A_4\rangle & \cdots\\
\vdots & \vdots &\vdots & \vdots & \vdots &\ddots
\end{array}\right].
\end{align} 
Now, the zero time-delay second-order correlation function for this state is
\begin{align}
g_j^{(2)}(0) &=\frac{\langle a_j^{\dagger 2} a_j^2 \rangle}{\langle a_j^\dagger a_j \rangle^2}=\frac{N_2\sum_{k=2} k(k-1)\langle A_k^\dagger A_k \rangle}{\left(\sum_{k=1} k\langle A_k^{\dagger} A_k \rangle \right)^2}, \nonumber\\
&=2^{(n-1)}g_p^{(2)}(0)\frac{N_2\left[1-\frac{(2+2^n)}{6}\eta^2 t^2-\left(\frac{(1+2^n)2^{n/2}+(3!)^{n/2}3^{n/2}}{2^{n/2}3!}-\frac{2(3!)^n}{2^{n}3!}\right)n_p\eta^2 t^2\frac{g_p^{(3)}(0)}{g_p^{(2)}(0)}+...\right]}{\left[1-\frac{1}{3}\eta^2 t^2+\frac{1}{36}\eta^4 t^4+\left(2^{(n-1)}-\frac{(1+2^n)}{3}\right)n_p \eta^2 t^2 g_p^{(2)}(0)+..... \right]^2}.
\end{align}
\section{Appendix B: n-photon single mode PDC process}\label{AppendixB}
The interaction Hamiltonian for this process is \cite{Drobny1994PRA} 
\begin{align}\label{Hamiltonian_degenerateNAppen}
\tilde{H}_{I}=\eta(a_{p}a_d^{\dagger n} +a_{p}^\dagger a_d^n).
\end{align}
The initial state at $t=0$ is given by $\ket{\psi}_p\ket{0}_d$, where $\ket{\psi}_p$ is the state of the pump field-mode and $\ket{0}_d$ is the vacuum of down-converted field-mode. Then the evolved state under the Hamiltonian given in Eq.~(\ref{Hamiltonian_degenerateNAppen}) at time $t$ is
\begin{align}
\ket{\Psi(t)}&=\frac{1}{\sqrt{N_4}}[B_0\ket{\psi}_p\ket{0}_d+B_1\ket{\psi}_p\ket{n}_d+B_2\ket{\psi}_p\ket{2n}_d+B_3\ket{\psi}_p\ket{3n}_d+B_4\ket{\psi}_p\ket{4n}_d+....],
\end{align} 
where
\begin{align*}
B_0 &=1-\left(\frac{n!\eta^2 t^2}{2}-\frac{(n!)^2\eta^4 t^4}{4!}\right) a_p^\dagger a_p+\left(\frac{(2n)!}{4!}+\frac{(n!)^2}{4!}\right) \eta^4 t^4 a_p^{\dagger 2}a_p^2+..,\\
B_1 &=i\sqrt{n!}\left[\left(-\eta t+\frac{n!\eta^3 t^3}{3!}\right) a_p+\frac{\eta^3 t^3}{3!}\left(\frac{(2n)!}{n!}+n!\right)a_p^\dagger a_p^2+...\right],\\
B_2 &=\frac{\sqrt{(2n)!}}{2!}\left[\left(-\eta^2 t^2+\left(\frac{n!}{6}+\frac{(2n)!}{n!}\right)\eta^4 t^4\right) a_p^2+\left(\frac{(3n)!}{(2n)!}+n!+\frac{(2n)!}{n!}\right)\frac{\eta^4t^4}{12}a_p^\dagger a_p^3+..\right],\\
B_3 &=i\frac{\sqrt{(3n)!}}{3!} \eta^3 t^3 a_p^3+..,\\
B_4 &=\frac{\sqrt{(4n)!}}{4!}\eta^4 t^4 a_p^4+.... 
\end{align*}
The state of the down-converted field-mode can be calculated by tracing over the pump state. Hence, the reduced density matrix of the down-converted field-mode in the basis $\{\ket{0}_d,\ket{n}_d,\ket{2n}_d,\ket{3n}_d,...\}$  is
\begin{align}\label{StateCollinearAppend}
\rho_{d}(t)&=\text{Tr}_{p} (\ket{\Psi(t)}\bra{\Psi(t)}),\nonumber\\
=\frac{1}{N_4}&\left[\begin{array}{cccccc}
\langle B_0^\dagger B_0\rangle & \langle B_0^\dagger B_1\rangle & \langle B_0^\dagger B_2\rangle & \langle B_0^\dagger B_3\rangle & \langle B_0^\dagger B_4\rangle & \cdots\\
\langle B_1^\dagger B_0\rangle & \langle B_1^\dagger B_1\rangle & \langle B_1^\dagger B_2\rangle & \langle B_1^\dagger B_3\rangle & \langle B_1^\dagger B_4\rangle & \cdots\\
\langle B_2^\dagger B_0\rangle & \langle B_2^\dagger B_1\rangle & \langle B_2^\dagger B_2\rangle & \langle B_2^\dagger B_3\rangle & \langle B_2^\dagger B_4\rangle & \cdots\\
\langle B_3^\dagger B_0\rangle & \langle B_3^\dagger B_1\rangle & \langle B_3^\dagger B_2\rangle & \langle B_3^\dagger B_3\rangle & \langle B_3^\dagger B_4\rangle & \cdots\\
\langle B_4^\dagger B_0\rangle & \langle B_4^\dagger B_1\rangle & \langle B_4^\dagger B_2\rangle & \langle B_4^\dagger B_3\rangle & \langle B_4^\dagger B_4\rangle & \cdots\\
\vdots & \vdots &\vdots & \vdots & \vdots &\ddots
\end{array}\right].
\end{align}
Here $\langle B_j^\dagger B_k \rangle=~_p\bra{\psi} B_j^\dagger B_k \ket{\psi}_p$.

The zero time-delay second-order correlation function for the above state is
\begin{align}
g_d^{(2)}(0)&=\frac{\langle a_d^{\dagger 2}a_d^2\rangle}{\langle a_d^\dagger a_d\rangle^2}=\frac{N_4\sum_{k=1}nk(nk-1)\langle B_k^\dagger B_k\rangle}{\left(\sum_{k=1}nk\langle B_k^\dagger B_k\rangle\right)^2},\nonumber\\
&=\frac{n-1}{n! n n_p \eta^2 t^2}\frac{N_4\left[\left(1-\frac{n!}{3}\eta^2 t^2\right)-\left[\frac{1}{3}\left(\frac{(2n)!}{n!}+n!\right)-\frac{2n-1}{2(n-1)}\frac{(2n)!}{n!}\right]n_p\eta^2 t^2
g_p^{(2)}(0)+..\right]}{\left[\left(1-\frac{n!}{3}\eta^2 t^2\right)-\left[\frac{1}{3}\left(\frac{(2n)!}{n!}+n!\right)-\frac{1}{2}\frac{(2n)!}{n!}\right]n_p\eta^2 t^2
g_p^{(2)}(0)+...\right]^2}.
\end{align}
\end{strip}
\section{Acknowledgment}
We acknowledge financial support through the research
grant no. EMR/2015/001931 from the Science and Engineering
Research Board (SERB), Department of Science
and Technology, Government of India and through
the research grant no. DST/ICPS/QuST/Theme-
1/2019 from the Department of Science and Technology,
Government of India. NM acknowledges IIT Kanpur for postdoctoral fellowship. 
\section*{Disclosures}
The authors declare no conflicts of interest.

%

\end{document}